\documentclass[twocolumn,showpacs,preprintnumbers,amsmath,amssymb]{revtex4}

\usepackage{graphicx}
\usepackage{dcolumn}
\usepackage{bm}
\usepackage{epsfig}
\usepackage{color}

\newcommand{\de}{{de}}
\newcommand{\s}{{\rm s}}

\def\fun#1#2{\lower3.6pt\vbox{\baselineskip0pt\lineskip.9pt
  \ialign{$\mathsurround=0pt#1\hfil##\hfil$\crcr#2\crcr\sim\crcr}}}
\def\simgt{\mathrel{\lower0.6ex\hbox{$\buildrel {\textstyle >}
 \over {\scriptstyle \sim}$}}}
\def\simlt{\mathrel{\lower0.6ex\hbox{$\buildrel {\textstyle <}
 \over {\scriptstyle \sim}$}}}

\input epsf

\newcommand{\mpcoh}{\,h^{-1}\,{\rm Mpc}}

\def\be{\begin{equation}}
\def\ee{\end{equation}}
\def\ba{\begin{eqnarray}}
\def\ea{\end{eqnarray}}

\newcommand{\om}{\omega_{\rm BD}}
\newcommand{\bfk}{\mbox{\boldmath$k$}}

\begin{document}

\preprint{}

\title{Theoretical Priors On Modified Growth Parametrisations}  

\author{$^1$Yong-Seon Song\email{yong-seon.song@port.ac.uk}, $^{1,2}$Lukas Hollenstein, $^1$Gabriela Caldera-Cabral and $^1$Kazuya Koyama}

\affiliation{$^1$Institute of Cosmology $\&$ Gravitation, Dennis Sciama Building,
University of Portsmouth, Portsmouth, PO1 3FX, United Kingdom\\
$^2$D\'epartement de Physique Th\'eorique, Universit\'e de Gen\`eve, 24 Quai Ernest Ansermet, 1211 Gen\`eve 4, Switzerland}

\date{\today}

\begin{abstract}
Next generation surveys will observe the large-scale structure of the Universe with unprecedented accuracy. This will enable us to test the relationships between matter over-densities, the curvature perturbation and the Newtonian potential. Any large-distance modification of gravity or exotic nature of dark energy modifies these relationships as compared to those predicted in the standard smooth dark energy model based on General Relativity. In linear theory of structure growth such modifications are often parameterised by virtue of two functions of space and time that enter the relation of the curvature perturbation to, first, the matter over- 
density, and second, the Newtonian potential. We investigate the predictions for these functions in Brans-Dicke theory, clustering dark energy models and interacting dark energy models. We find that each theory has a distinct path in the parameter space of modified growth. Understanding these theoretical priors on the parameterisations of modified growth is essential to reveal the nature of cosmic acceleration with the help of upcoming observations of structure formation.
\end{abstract}

\pacs{draft}

\keywords{large-scale structure, structure formation, modified gravity, dark energy, theoretical cosmology}

\maketitle

\section{Introduction}

Our understanding of the laws of physics we gained on Earth has already been challenged by cosmological observations of the formation of galaxies and the accelerated expansion of the Universe \cite{Perlmutter:1998np, Riess:1998cb}. In order to establish a cosmological standard model without inconsistencies between observations and known physics, it is unavoidable either to assume the presence of as yet undetected dark materials filling 95$\%$ of the energy of the Universe, or to modify some of the fundamental laws of physics such as Einstein's General Relativity. The information on the underlying physics of the Universe is provided by looking at the formation of the large scale structure (LSS) of matter in the Universe. While current observations have revealed the breakdown of our knowledge of physics on cosmological scales, the future observation of the LSS will provide a clue to reveal the nature of the late-time cosmic acceleration; which part of our physical picture should be modified - matter or gravity, and how it is modified \cite{Ishak:2005zs, Song:2005gm, Wang:2007fsa, Knox:2006fh, Mortonson:2008qy, Linder:2005in, Linder:2007hg, Huterer:2006mva, Polarski:2007rr, DiPorto:2007ym, Thomas:2008tp, Dent:2008ia, Uzan:2000mz, Song:2006sa, Bertschinger:2006aw, Koivisto:2005mm, Amendola:2007rr, Uzan:2006mf, Hu:2007pj, Hu:2008zd, Zhao:2008bn, Bertschinger:2008zb, Daniel:2008et, Caldwell:2007cw, Dore:2007jh, Jain:2007yk, Song:2008vm, Zhang:2007nk, Zhang:2008ba, Song:2008xd}.

The current standard cosmological model was established based upon unknown dark materials and the validity of General Relativity (GR) on cosmological scales. We call this standard model \emph{smooth dark energy} (sDE). It consists of the following energy components: 5$\%$ baryons as described by the standard model of particle physics and 95$\%$ of unobserved dark particles -- 20$\%$ of collisionless dark matter, fitting galaxy formation and the peaks in the spectrum of cosmic microwave background (CMB) anisotropies, and 75$\%$ dark energy explaining the cosmic acceleration, described by a perfect fluid that is homogeneous (smooth) on sub-horizon scales. It is difficult to accept dark energy as a standard building block of energy components of the Universe as it is not yet supported neither phenomenologically nor theoretically. Gravity on cosmological scales can be modified to accelerate the cosmic expansion without dark energy. The degeneracy between GR and modified gravity (MG) models in the background expansion history is broken by the growth history of structure formation. Any departure from the growth history in sDE is probed by two functions expressing new degrees of freedom: the mass screening, $Q$, changes the relationship between the curvature perturbation and the matter over-densities, while the effective anisotropic stress, $\eta$, alters the relationship between the curvature perturbation and the Newtonian potential.

Here we study whether the possible detection of non-trivial $Q$ and $\eta$ can be considered as a probe of the break-down of the sDE model at cosmological scales. We show that non-trivial $Q$ and $\eta$ functions are also induced in other GR models via clustering of dark energy or interactions between dark energy and dark matter. GR will be confirmed (or excluded) by looking at the relation between $Q$ and $\eta$, not merely by a departure of $Q$ and/or $\eta$ from unity. We find that both MG models \cite{Dvali:2000hr, Carroll:2003wy} and exotic GR models \cite{Amendola:1999er} provide unique distinguishable trajectories that are well presented in the plane of two transformed parameters: $\Sigma\equiv Q(1+1/\eta)/2$ parameterises the relation between the lensing potential and matter over-densities and $\mu\equiv Q/\eta$ represents the relation between the Newtonian potential and the matter over-densities.
We examine these parameters in various models and show that each theory has a distinct path on the plane of those parameters. The understanding of these theoretical priors is essential to reveal the nature of the cosmic acceleration from future structure formation observations.

This paper is organised as follows. In section II  we introduce the {\it smooth dark energy} (sDE) model and the modified growth parameters $Q$ and $\eta$. 
Then we investigate these functions for different cases: In section III the Brans-Dicke (BD) theory is used to characterise modified gravity models on sub-horizon scales. In section IV we consider clustering dark energy (cDE) models and in section V interacting dark energy (IDE) models are studied. In section VI we show that these models have distinct paths in the plane of the modified growth parameters $\mu$ and $\Sigma$. We conclude in section VII.

\section{Parameterisation}

Any deviation from our standard cosmological model sDE will lead to modifications of the background evolution and structure growth. Let us fix the background using the sDE reference model and, subsequently, parameterise the modified growth in terms of the functions $Q$ and $\eta$ such that these become trivial ($Q=\eta=1$) in sDE.

We assume a flat Friedmann-Robertson-Walker (FRW) background metric. In the sDE model the Friedmann equation defines the evolution of the Hubble parameter, $H$, in terms of the energy content of the Universe
\be
  H^2 = \frac{8 \pi G_N}{3} \left(\rho_m^\s + \rho_\de^\s\right) \,.
  \label{eq:FRGR}
\ee
We introduced the label ``$\s$'' to mark the quantities given in the sDE model. In any other model we adjust the parameters such that $H$ evolves in the same way as in the sDE reference model.

Next, we consider linear perturbations around the flat FRW background described in the Newtonian gauge:
\be
  ds^2 = -(1+2\Psi)dt^2 + a^2(1+2\Phi)\delta_{ij}dx^idx^j \,.
\ee
We will use a dot to denote the derivative with respect to cosmic time, $t$. In GR the Newtonian potential $\Psi$ and the curvature perturbation $\Phi$ are related to the matter content via the linearised Einstein equations. In sDE these yield two constraint equations. Firstly, because there is no anisotropic stress at late times the two metric potentials coincide, $\Psi^\s=-\Phi^\s$, and secondly, the Poisson equation relates $\Phi$ to the comoving matter density perturbation $\Delta_m$:
\be
  k^2 \Phi^\s = 4 \pi G_N a^2 \rho_m^\s \Delta_m^\s \,.
\ee
For a general fluid ($i$) the comoving density perturbation is defined as
\be
  \Delta_i \equiv \delta_i -\frac{\dot\rho_i}{\rho_i}\frac{\theta_i}{k^2} \,,
\ee
in terms of the density contrast $\delta_i$ and the velocity divergence $\theta_i$ in the Newtonian gauge. The comoving density perturbation $\Delta_m$ is useful as it is the over-density measured by observers comoving with the matter-flow.

In any other GR dark energy or MG model the Poisson equation is modified. This departure from sDE in the Poisson equation is parameterised by the function $Q(a,k)$ such that
\ba
  k^2 \Phi = 4 \pi G_N a^2 Q(a,k) \rho^\s_m \Delta_m \,,
  \label{def:Q}
\ea
holds with $\Delta_m$ being the comoving matter perturbation as measured in the modified model (with the same background evolution as sDE) and $\rho_m^\s$ that scales in the usual way with time ($\propto a^{-3}$). Because $\Phi$ and $\Delta_m$ are gauge-invariant variables this is a gauge-invariant definition of $Q$. In MG models, $Q$ represents the mass screening effect due to a local modification of gravity, i.e.~an effective modification of Newton's constant. In exotic dark energy models within GR it measures the additional clustering due to a modification of the background evolution of the energy densities and different evolution of the perturbations.

In order to fully describe the  modified growth of perturbations, the other constraint equation of GR needs to be parameterised as well. The relation between $\Phi$ and $\Psi$ is altered such that
\be
  \eta \equiv -\frac{\Phi}{\Psi}
\ee
is not necessarily trivial. As compared to sDE in GR, where $\Phi/\Psi=-1$, a non-trivial $\eta$ effectively adds an anisotropic stress component at linear level.

Measuring $\Phi$ and therefore $Q$ directly from the galaxy distribution is difficult due to galaxy bias. Therefore, we will also present our results in terms of two re-defined functions that are more directly connected to unbiased observables. First, we notice that weak lensing and the late ISW effect are sensitive to $(\Phi-\Psi)/2$ and therefore it is convenient to use the combination
\be
  \Sigma \equiv \frac{Q}{2}\left(1+\frac{1}{\eta}\right) \,,
  \label{def:Sigma}
\ee
while peculiar velocities are responding to $\Psi$ and, thus, they determine
\ba
  \mu \equiv \frac{Q}{\eta} \,.
  \label{def:mu}
\ea
Note that the growth of matter perturbations is also determined by the Newtonian potential, and hence by $\mu$. Therefore, these transformed functions are more closely related to observations.

In the following sections we will study the evolution of matter fluctuations inside the Hubble horizon. Then we can use the quasi-static approximation and neglect the time derivatives of the perturbed quantities compared with the spatial derivatives. Furthermore, the comoving density perturbation and the Newtonian gauge density contrast are approximately equal, $\Delta_i\simeq\delta_i$, which means we can neglect the contribution from the velocity potentials to the Poisson equation. However, this is not a perfect approximation when considering dark energy perturbations as we will discuss in Sec~\ref{sec:cde}.

\section{Quasi-static perturbations in modified gravity models}
  \label{sec:quasi-static_PT}

As mentioned in the introduction, the large distance modification of gravity, which is necessary to explain the late-time acceleration, generally modifies gravity even on sub-horizon scales due to the introduction of a new scalar degree of freedom. This modification of gravity due to the scalar mode can be described by Brans-Dicke (BD) gravity \cite{Bertschinger:2008zb, Zhao:2008bn, Koyama:2009me}. In the quasi-static approximation the perturbed modified Einstein equations give
\begin{eqnarray}
\Phi + \Psi &=&  \varphi,
\label{eq1}
\\
k^2 \Phi &=& 4 \pi G_Na^2 \rho_m \delta_m + \frac{k^2}{2} \varphi,
\label{eq:Poisson_eq}
\\
(3 +2 \omega_{\rm BD}) k^2 \varphi
&=& - 8 \pi G_Na^2 \rho_m \delta_m,
\end{eqnarray}
where $\rho_m$ is the background dark matter energy density, $\delta_m$ is dark matter density contrast and $\varphi$ is the BD scalar field. In the quasi-static approximation the BD parameter $\omega_{\rm BD}$ and Newton's constant $G_N$ can be any functions of time. In the following we assume that $G_N$ can be approximated by the one measured in Cavendish-like experiments and the it is constant.

In general, modified gravity models that explain the late-time acceleration predict $\omega_{\rm BD} \sim O(1)$ on sub-horizon scales today. This would contradict the solar system constraints which require $\omega_{\rm BD} > 40000$. However, this constraint can be applied only if the BD scalar has no potential and no self-interactions. Thus, in order to avoid this constraint, the BD scalar should acquire some interaction terms on small scales. Generally, we expect that the BD scalar field equation is given by \cite{Koyama:2009me}
\begin{equation}
(3 +2 \omega_{\rm BD}) k^2 \varphi
= -8 \pi G_Na^2 \rho_m \delta_m - {\cal I}(\varphi) \,,
\label{eq:BD_eq}
\end{equation}
in a Fourier space. Here the interaction term ${\cal I}$ can be expanded as
\begin{eqnarray}
{\cal I} (\varphi)
&=& M^2 \varphi(k) + \frac{1}{2} \int \frac{d^3 \bfk_1 d^3 \bfk_2}
{(2 \pi)^3} \delta_D(\bfk -\bfk_{12}) \times
\nonumber \\
&& \times
M_2(\bfk_1, \bfk_2) \varphi(\bfk_1) \varphi(\bfk_2) + \ldots
\label{eq:BD_eq2}
\end{eqnarray}
where $\bfk_{ij}=\bfk_i+\bfk_j$.  Non-linear terms become important when considering non-linear clustering of dark matter and they are crucial to recover GR on non-linear scales. However, in this work we are interested in linear perturbations thus we only take into account the linear term.

From Eqs.~(\ref{eq1})-(\ref{eq:BD_eq2}) we can directly derive the functions $Q(a,k)$ and $\eta(a,k)$:
\begin{eqnarray}\label{eq:MGqeta}
Q(k,a) &=& \frac{2(1+ \om)+ M^2 a^2/k^2}{3 +2 \om + M^2 a^2/k^2} \,, \\
\eta(k,a) &=& \frac{2(1+\om) + M^2 a^2/k^2}{2(2+\om) + M^2 a^2/k^2} \,.
\end{eqnarray}
There are two ways to recover GR. One is to take $|\om| \gg 1$ and the other one is to consider large scales $Ma/k \gg 1$. In both cases $Q$ and $\eta$ approach unity. For $M=0$, there is a unique trajectory in the $Q-\eta$ plane parameterised by $\om$.

As discussed in the previous section, the parameters that are more closely related to observations are $\Sigma$ and $\mu$; see the definitions (\ref{def:Sigma}) and (\ref{def:mu}).
We find that these parameters are given by
\begin{equation}
\Sigma(k,a)=1, \quad \mu(k,a) = \frac{2(2+\om) + M^2 a^2/k^2}{3 + 2 \om +M^2 a^2/k^2} \,.
\end{equation}
Interestingly, $\Sigma(k,a)=1$ is the same as in GR. This is because the scalar field couples to the trace of the energy-momentum tensor and it does not couple to the photon. Since $\Sigma$ parameterises the lensing potential, which in turn determines the geodesics of photons, it is not affected by the scalar field and, thus, $\Sigma=1$ as in GR.

On the other hand, $\mu$ measures the strength of Newton gravity.
In order to understand the result, it is instructive to find a solution for the scalar filed $\varphi$ with a localised matter source $\delta \rho_m$ in the Mikowski background. Let us first consider the case with $M=0$. Assuming spherical symmetry, the solution for $\varphi(r)$ far away from the source in real space is given by
\begin{equation}
\varphi(r) = - \frac{1}{3 + 2 \om} \frac{r_g}{r} \,,
\end{equation}
where $r_g \equiv 8 \pi G_N \int dr' r'^2 \delta \rho_m$ is the gravitational length of the source. If  $3 + 2 \om >0$, the scalar field mediates an attractive force. From Eqs.~(\ref{eq1}) and (\ref{eq:Poisson_eq}) we obtain an equation for the Newtonian potential
\begin{equation}
k^2 \Psi =  - 4 \pi G_Na^2 \rho_m \delta_m + \frac{k^2}{2} \varphi \,.
\end{equation}
Then we can see that the scalar field adds an additional attractive force to the Newtonian potential if $3 + 2 \om >0$ and, consequently, $\mu$ becomes larger than one. 
On the other hand, for $3 + 2 \om <0$ the scalar field mediates a repulsive force. This is related to the fact that for $3 + 2 \om <0$ the scalar field becomes a ghost that has a kinetic term with the wrong sign. In this case, the scalar field suppresses the Newtonian force and $\mu$ becomes smaller than one.

Next, let us discuss the effect of the mass. For $3 + 2 \om >0$ the solution for the scalar field in case of a spherically symmetric source becomes
\begin{equation}
\varphi(r) = - \frac{1}{3 + 2 \om} \frac{r_g}{r}
\exp \left[ - \frac{r}{r_M} \right] \,,
\end{equation}
where
\begin{equation}
r_M \equiv \sqrt{\frac{3+2\om}{M^2}}
\end{equation}
is the Compton wavelength of the scalar field.
On large scales, $r > r_M$, the solution exponentially decays and the scalar field becomes negligible. In this case there is no scalar force and GR is recovered, so $\mu =1$. On the other hand, for $r < r_M$, the effect of the mass can be neglected and we get the same result as for the massless case. Thus, for $M \neq 0$, $\mu$ becomes scale-dependent.

This BD theory contains two well known modified gravity models as special cases: $f(R)$ gravity and the DGP braneworld model.  Firstly, in $f(R)$ gravity we have $\om=0$ \cite{Carroll:2003wy}. In this case $\mu = 4/3$ for $Ma/k \ll 1$ and $\mu =1$ for $Ma/k \gg 1$. In $f(R)$, though, $G_N$ can be time-dependent, which effectively modifies $\Sigma$ and $\mu$. However, its dependence on time is strongly constrained by observations and we can safely neglect such an effect.
Secondly, in the DGP model we have $\om = (3/2)(\beta(t) -1)$ with $\beta(t)= 1- 2 H r_c (1 + \dot{H}/3 H^2)$, where $\dot{H}$ is the time derivative of the Hubble parameter and $r_c \sim H_0^{-1}$ is the cross-over scale, which is a parameter in this model \cite{Dvali:2000hr}. Here $M=0$ for quasi-static perturbations. We should note that $3 +2 \om <0$ and $\mu <1$ in DGP indicating a theoretical pathology of the model \cite{Koyama:2007za}.


\section{Clustering dark energy}  \label{sec:cde}

The dark energy can in principle support long-lived fluctuations \cite{Bean:2003fb, Weller:2003hw, Hu:2004yd, Takada:2006xs, Avelino:2008cu, Unnikrishnan:2008qe, Abramo:2009ne, Battye:2009ze, Sapone:2009mb}. In these clustering dark energy (cDE) models the baryons will fall into the potential wells created by both the dark matter and energy. Here we consider a general dark energy fluid with a time-varying barotropic equation of state parameter, $w_\de\equiv P_\de/\rho_\de$, and some scalar anisotropic stress $\sigma_\de$. However, in this section we restrict the dark energy to be minimally coupled to gravity and (dark and baryonic) matter, which means that here $\rho_i=\rho_i^s$ for $i=m,\,\de$; and we can treat the baryons and the dark matter as a single pressureless perfect fluid.

Given the different physical behaviour of matter and dark energy the total density fluctuation will not be a simple linear function of $\delta_m$ but acquires a contribution from the dark energy. The Poisson equation in cDE models is then given by
\be
  k^2\Phi = 4\pi G_N a^2 (\rho_m\delta_m+\rho_\de\delta_\de)
\ee
The dark energy anisotropic stress, $\sigma_\de$, enters the second Einstein constraint equation as
\be
  k^2(\Phi+\Psi)=-12\pi G_Na^2(1+w_\de)\rho_\de\sigma_\de \,.
\ee
Because the evolution of the background energy densities in cDE is not modified with respect to the sDE reference, we have $\rho_m^\s=\rho_m$. So the definition of Q in Eq.~(\ref{def:Q}) directly implies
\be
Q(a,k)=1+\frac{\rho_\de\delta_\de}{\rho_m\delta_m}\,,
\label{cde_Q}
\ee
for cDE models. By combining the two constraints we find that the $\eta$ parameter here is
\be
  \eta=\frac{\rho_m\delta_m+\rho_\de\delta_\de}
    {\rho_m\delta_m+\rho_\de\delta_\de+3(1+w_\de)\rho_\de\sigma_\de} \,.
  \label{cde_eta}
\ee

As we want to get a feeling for the time- and scale-dependence of $Q$ and $\eta$ we take a look at the evolution equations of $\delta_m$ and $\delta_\de$ without specifying the dark energy pressure and anisotropic stress perturbations. Later we will consider a simple model  for the pressure perturbation and a specific model for the anisotropic stress of dark energy motivated by the possibility of having scale-invariant growth of dark energy perturbations.

From the energy-momentum conservation of the matter and the dark energy we can derive second-order differential equations for $\delta_m$ and $\delta_\de$ in the quasi-static approximation where time derivatives of the gravitational potentials are neglected. For the matter component we find the well-known growth equation
\be
  \ddot{\delta}_m +2H\dot\delta_m = -\dfrac{k^2}{a^2}\Psi \,,
\ee
with the source term as found from the two Einstein constraint equations above:
\be
  -\frac{k^2}{a^2}\Psi = \frac{3}{2}H^2\left\{\Omega_m\delta_m
    + \Omega_\de\delta_\de +3(1+w_\de)\Omega_\de\sigma_\de \right\} \,.
  \label{cde_source}
\ee
Here $\Omega_i=\Omega_i(t)$ are the fractional energy densities as functions of time.
For the general barotropic dark energy fluid perturbations we find the evolution equation
\ba
  && \hspace{-3mm} \ddot{\delta}_{de} +(2-6w_\de)H\dot\delta_{de}
    +3H\left(\!\dfrac{\delta P_\de}{\rho_\de}\!\right)^{\!\!\text{\large .}} =
    \nonumber \\
  && \hspace{2mm} 3H\dot{w}_\de\delta_\de + 3\left[(2-3w_\de)H^2+\dot{H}\right]
    \left[w_\de\delta_\de -\dfrac{\delta P_\de}{\rho_\de}\right]
  \nonumber \\
  && \hspace{2mm} -(1+w_\de)\dfrac{k^2}{a^2}\left[\dfrac{\delta P_\de}{(1+w_\de)\rho_\de}
      -\sigma_\de +\Psi\right] \,.
  \label{cde_degrowth}
\ea
\emph{A priori} there are no evolution equations for $\delta P_\de$ and $\sigma_\de$ available. To close the system we need some relation between the pressure and anisotropic stress perturbations and the density perturbation.

Formally, cDE can predict any $Q(a,k)$ and $\eta(a, k)$. From Eqs.~(\ref{cde_Q}) and (\ref{cde_eta}), we can derive $\delta_{de}$ and $\sigma_{de}$ in terms of $Q$ and $\eta$. Then substituting them into Eq.~(\ref{cde_degrowth}), we can find a solution for $\delta P_{de}$. However, in physical models, $\delta P_\de$ and $\sigma_\de$ should be determined by microphysics of the dark energy and it is very unlikely that we have enough degrees of freedom to fine-tune these quantities to get any desirable $Q$ and $\eta$. Of course it is a formidable task to find a theoretical prior for dark energy clustering. In the following, we consider a toy model for $\delta P_\de$ and $\sigma_\de$ to get some idea.

\begin{figure}[t]
\centering
\includegraphics[width=\columnwidth]{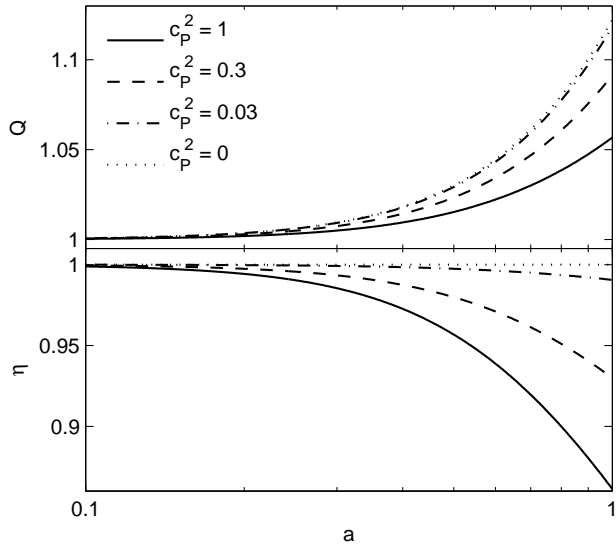}
\caption{The evolution of $Q$ and $\eta$ is presented with varying $c_P^2$ for the case of scale-invariant growth of dark energy perturbations where the anisotropic stress balances the pressure support ($f_{\sigma}=1$). Here we set $w_\de=-0.8$ as in the sDE reference.}
\label{cde_scaleinv_Qeta_evol}
\end{figure}

The pressure perturbation is related to the density perturbation via the speed of sound. Usually one defines the speed of sound in the rest-frame of the fluid where the pressure perturbations are assumed to be adiabatic: $\hat c_s^2\equiv(\delta{}P_{de}/\delta\rho_\de)_{\rm rest}$. The transformation back into the Newtonian gauge leads to a correction to $\delta{}P_{de}$ which is proportional to the velocity potential, $\theta_{de}/k^2$.
Here we simply parameterise the pressure perturbation as
\be
  \frac{\delta{}P_\de}{\rho_\de} \equiv c_P^2\delta_\de +\frac{g_P}{k^2} \,.
\label{dpressure}
\ee
Both, $c_P^2$ and $g_P$ can be functions of time and space. However, we will consider $c_P^2$ to be constant
\footnote{The definition of
$c_P^2$ would coincide with the conventional rest-frame sound speed, $\hat c_s^2$, if $g=3H(1+w)(\hat c_s^2 -c_a^2)\theta_\de$ where the adiabatic sound speed is $c_a^2\equiv \dot P_\de/\dot\rho_\de$. See W.~Hu, Astrophys.\ J.\  {\bf 506}, 485 (1998), astro-ph/9801234}.

In the absence of anisotropic stress, $\sigma_\de=0$, dark energy perturbations are strongly suppressed on comoving scales smaller than $c_P/(aH)$ because of the negative scale-dependent term $\propto k^2\delta P_\de$ on the rhs of the evolution equation (\ref{cde_degrowth}). Therefore, dark energy perturbations become only important on quasi-static scales if $c_P^2\ll 1$.

The presence of anisotropic stress in the dark energy, however, can mitigate this effect by compensating the pressure support. The extreme case where $\sigma_\de = \sigma^0_\de$
with
\be
  \sigma^0_\de \equiv \frac{\delta P_\de}{(1+w_\de)\rho_\de} \,,
\ee
leads to a scale-independent growth of dark energy perturbations if $g_P=0$. In this case the scale-dependent suppression term on the rhs of the evolution equation for $\delta_\de$ vanishes. More generally we set
\be
\sigma_\de = f_{\sigma}\sigma^0_\de+\frac{g_{\sigma}}{k^2} \,,
\ee
such that the constant parameter $f_\sigma$ re-introduces scale-dependence into the growth of perturbations in case $f_\sigma\neq 1$. This effect becomes more apparent when we substitute the expressions for the pressure and anisotropic stress perturbations into the evolution equation. For the simple case when $w_\de$ and $c_P^2$ are constants and $g_P=g_\sigma=0$ we find from Eq.~(\ref{cde_degrowth})
\ba
  && \hspace{-5mm} \ddot{\delta}_{de} +(2+3c_P^2-6w_\de)H\dot\delta_{de} =
    \nonumber \\
  && \hspace{5mm} \frac{3}{2}(1+w_\de)H^2\Big\{ \Omega_m\delta_m +
      (1+3f_\sigma c_P^2)\Omega_\de\delta_\de
    \nonumber \\
  && \hspace{5mm} +2\frac{w_\de-c_P^2}{1+w_\de}\Big(2-3w_\de+\frac{d\ln H}{d\ln a}\Big)
        \delta_\de \Big\}
  \nonumber \\
  && \hspace{5mm} -\dfrac{k^2}{a^2}(1-f_\sigma)c_P^2\delta_\de
  \label{cde_degrowth_simple}
\ea
where we also substituted the source term from Eq.~(\ref{cde_source}). For the matter we find
\be
  \ddot{\delta}_m +2H\dot\delta_m = \frac{3}{2}H^2\Big\{\Omega_m\delta_m
    + (1+3f_\sigma c_P^2)\Omega_\de\delta_\de \Big\} \,.
  \label{cde_mgrowth_simple}
\ee
In the following we will investigate the two cases: scale-independent growth of dark energy perturbations under the conditions $f_\sigma=1$, $g_P=0$ and $g_\sigma=0$, and scale-dependent growth with a deviation from either one of those conditions.

On scales where the quasi-static approximation is applicable, the clustering of dark energy is governed by the competition between the gravitational infall and the pressure support as can be seen from the $k^2$-dependent term in the full evolution equation (\ref{cde_degrowth}). However, the anisotropic stress counteracts the pressure. For $\sigma_\de=\sigma_\de^0$, i.e.~$f_\sigma=1$, the anisotropic stress exactly cancels the pressure support. The remaining source term is dominated by the matter density. Thus, the dark energy clusters on all scales like dark matter does, and consequently, the growth of perturbations becomes scale-independent.

From the simplified evolution equation (\ref{cde_degrowth_simple}), with constant $w_\de$ and $c_P^2$, we can read off the influence of the sound speed: $c_P^2>0$ enhances the Hubble drag, the second term on the lhs of Eq.~(\ref{cde_degrowth_simple}), and also lowers the source term, as the third term in the braces is $<-10$ for typical values of $w_\de$ and all $c_P^2>0$. In other words, compared to the matter perturbations, dark energy perturbations grow weakly compared with dark matter perturbations but are not completely suppressed. Therefore we expect $Q>1$ at late times. Dark matter perturbations do not suffer the suppression but are rather enhanced due to the dark energy anisotropic stress adding to the source term, as seen from Eq.~(\ref{cde_mgrowth_simple}). The clustering condition of $c_P^2\ll 1$ is relaxed compared to the case with no anisotropic stress and we find dark energy perturbations cluster on all scales even for $c_P^2=1$. In Fig.~\ref{cde_scaleinv_Qeta_evol} we show the late-time evolution of $Q$ and $\eta$ in the case of the scale-independent growth. With the help of a modified version of the CAMB code \cite{Lewis:1999bs} we plot the mode $k=0.01\mpcoh$, which is representative for the quasi-static regime. It is important to note that when computing quantities like $Q$ with the help of a full Boltzmann-code like CAMB rather than within the quasi-static approximation it is necessary to use the correct gauge-invariant definition of $Q$ in terms of the comoving density perturbations $\Delta_i$ rather than the Newtonian gauge quantities $\delta_i$. Omitting this leads to artificial scale-dependence because it takes time for modes to relax to their late-time asymptotic behaviour after entering the horizon.

\begin{figure}[t]
\centering
\includegraphics[width=\columnwidth]{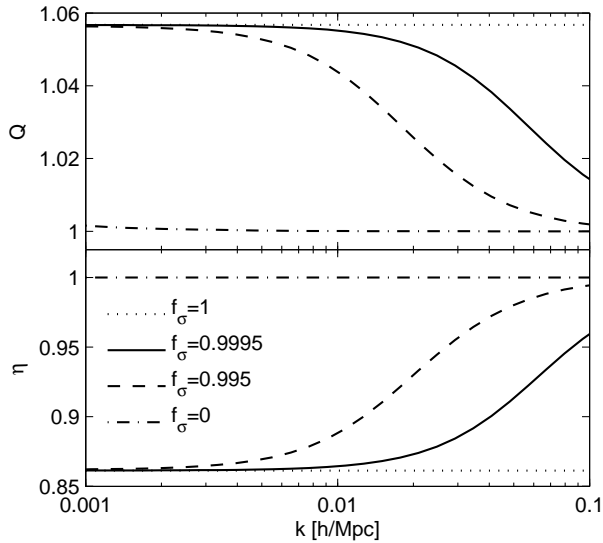}
\caption{The scale-dependence of $Q$ and $\eta$ is shown for different values of $f_\sigma$. We include the limiting cases of no anisotropic stress ($f_\sigma=0$, dash-dotted) and scale-independent growth of dark energy perturbations ($f_\sigma=1$, dotted). Here we set $w_\de=-0.8$ and $c_P^2=1$ as in the sDE reference.}
\label{cde_scaledep_Qeta_k}
\end{figure}

$Q$ is enhanced for smaller $c_P^2$. The anisotropic stress that is needed to cancel the pressure support, $\sigma_\de=\sigma_\de^0$, leads to $\eta<1$ as can be directly seen from Eq.~(\ref{cde_eta}). If $c_P^2=0$ there is no anisotropic stress, but also there is no scale-dependent suppression of the source term in the growth equation for dark energy perturbations Eq.~(\ref{cde_degrowth_simple}). Therefore dark energy perturbations cluster on all scales like dark matter does. Though the growth of dark energy perturbations is suppressed compared with dark matter perturbations because of a larger Hubble drag and a scale-independent suppression of the source term.

If the anisotropic stress does not exactly cancel the scale-dependent pressure support in the dark energy growth equation (\ref{cde_degrowth}), i.e.~$f_\sigma\neq 1$, then the growth becomes scale-dependent. The term $\propto k^2$ on the rhs of the dark energy growth equation becomes important on quasi-static scales. For $f_\sigma>1$ it acts as an instability and $\delta_{de}$ would blow up on small scales. For $f_\sigma<1$ we can define the scale
\be
  \lambda_\sigma \equiv \frac{2\pi c_P\sqrt{1-f_\sigma}}{aH}
\ee
which acts as a reduced sound horizon such that dark energy only clusters on scales larger than $\lambda_\sigma$ (as opposed to the scale-invariant case where dark energy clusters on all scales equally.)  A slight departure of $f_\sigma$ from unity therefore effectively decreases the sound horizon as compared to the sDE. Therefore we expect dark energy to cluster on smaller scales than in the case where there is no anisotropic stress. As a result, $Q$ and $\eta$ become scale-dependent and do approach unity on scales smaller than $\lambda_\sigma$, see Fig.~\ref{cde_scaledep_Qeta_k}. In addition to this case, scale-dependent $Q$ and $\eta$ are generated by $g_P\ne 0$ or $g_\sigma\ne 0$, as the scale-dependent effects from the pressure perturbation or anisotropic stress cause the density contrast to evolve in a scale-dependent way. We leave the investigation of these effects for future work.

\section{Interacting dark energy}

Dark matter is currently only detected via its gravitational
effects, and there is an unavoidable degeneracy between dark
matter and dark energy within GR. There could be a
hidden non-gravitational coupling between dark matter and dark
energy. In these interacting dark energy (IDE) models the
continuity equations for cold dark matter and dark energy are
modified by the coupling $\mathcal{C}$ such that
\begin{eqnarray}
\dot\rho_c+3H\rho_c&=&-\mathcal{C} \,,
\label{rhocb}\\
\dot\rho_\de+3H(1+w_\de)\rho_\de&=&\mathcal{C} \,.
\label{rhoxb}
\end{eqnarray}
where $w_\de=P_\de/\rho_\de$. $\mathcal{C}$ is the rate of the energy transfer, so that
$\mathcal{C}>0(<0)$ implies that energy is transferred from dark matter to dark energy (from dark energy to dark matter). In the following, we consider two types of interactions and identify $Q$ and $\eta$ in these models. In this section, we ignore the clustering of dark energy studied in the previous section assuming that the sound speed is one and there is no anisotropic stress. Also we assume baryons are not coupled to dark energy as there are strong constraints from laboratory tests.

\begin{figure*}[ht]
\centering
\includegraphics[width=0.45\textwidth]{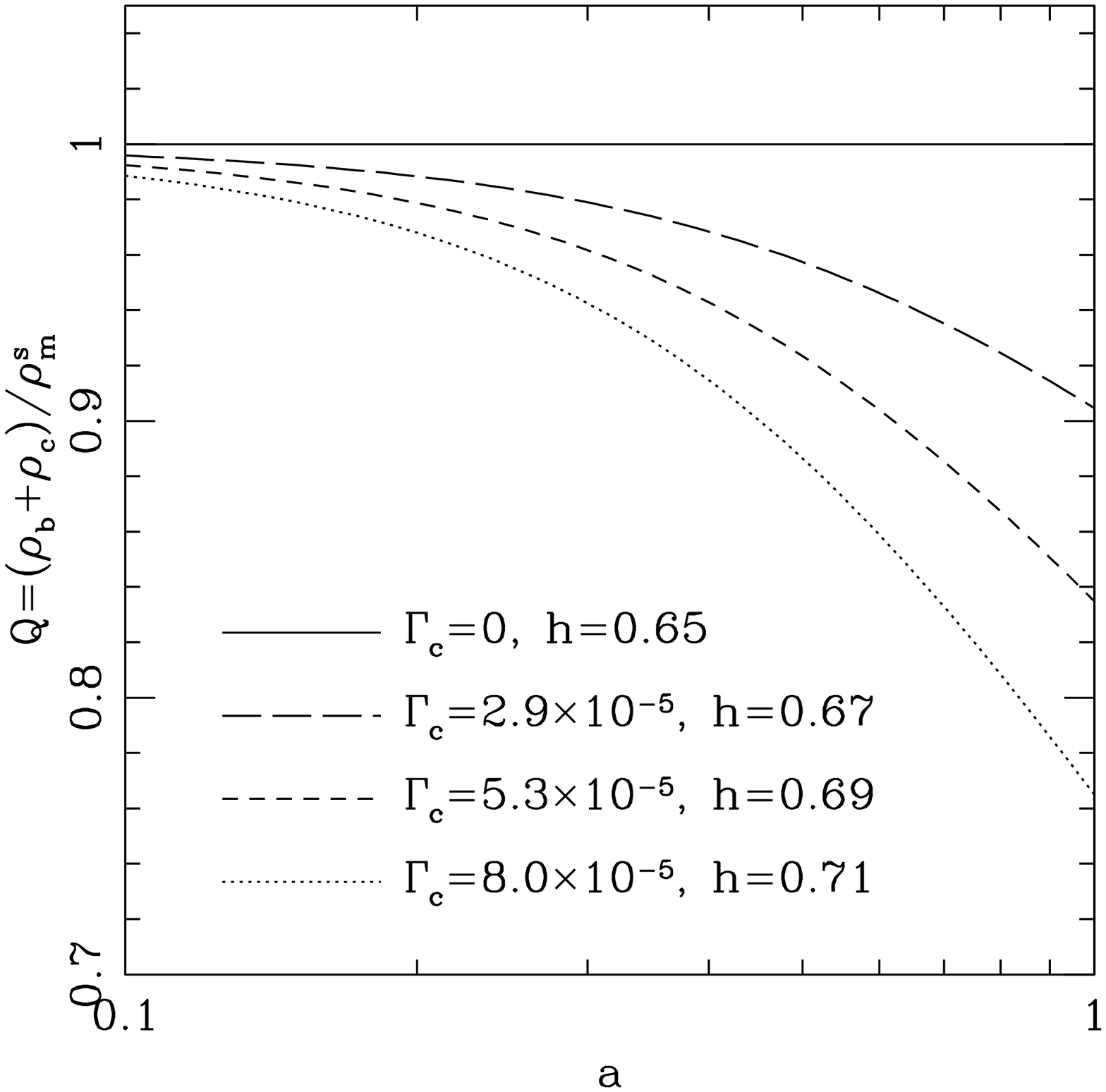}\hfill
\includegraphics[width=0.45\textwidth]{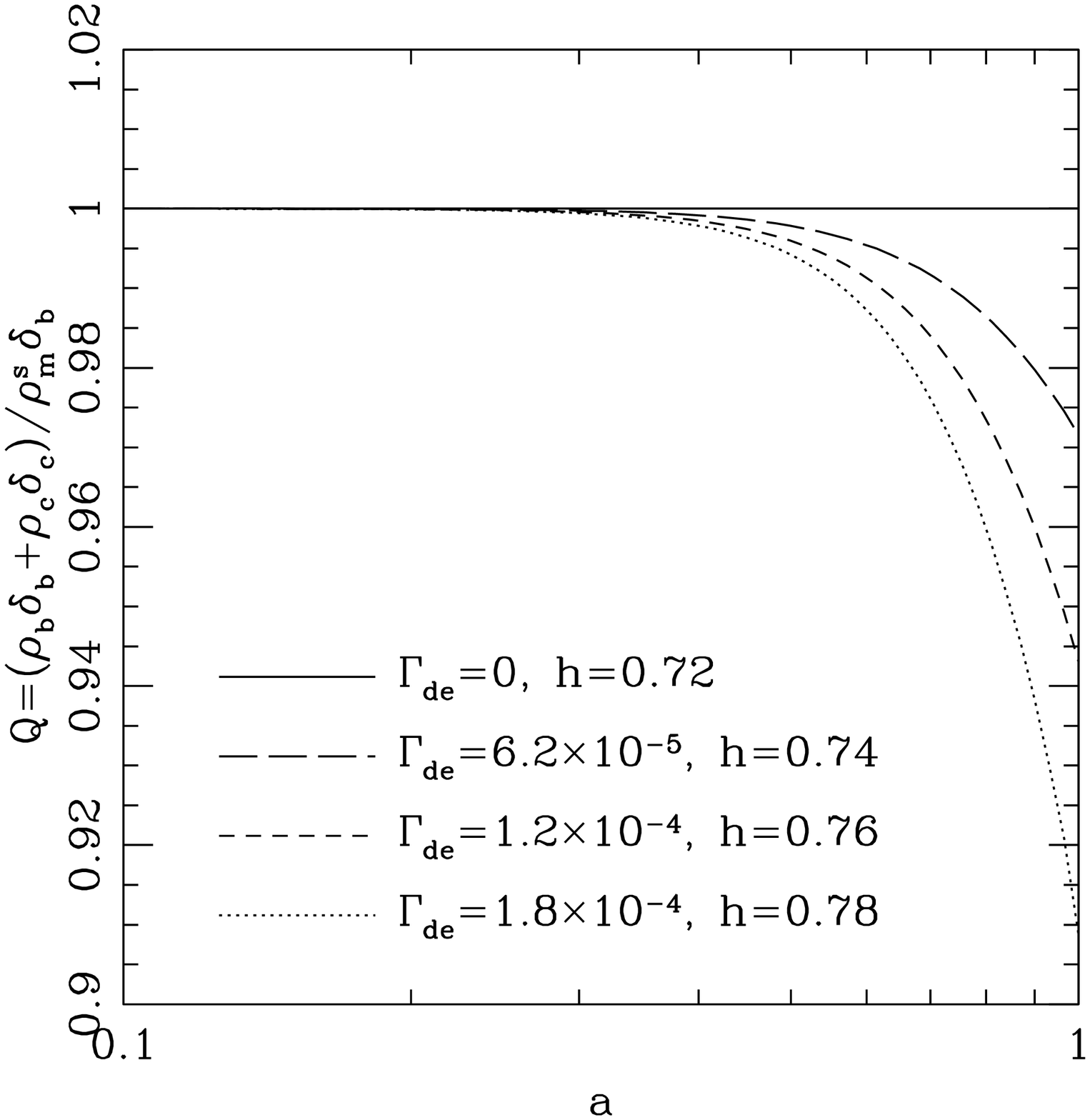}
\caption{The time evolution of $Q$ is presented for two different models of IDE. The left-hand side plot corresponds to Model I: $\mathcal{C}=\Gamma_c\rho_c$ with varying $\Gamma_c$ (1/Mpc). The right-hand side plot corresponds to Model II: $\mathcal{C}=\Gamma_\de\rho_\de$ with varying $\Gamma_\de$ (1/Mpc).}
\label{qcx}
\end{figure*}

\subsection{Model I: $\mathcal{C}=\Gamma_c\rho_c$}

First we consider the interaction given by $\mathcal{C}=\Gamma_c\rho_c$. A covariant form of the interaction can be found in \cite{CalderaCabral:2008bx, Valiviita:2008iv, CalderaCabral:2009ja} where it was shown that for this form of interaction there are no modifications to the dynamics of quasi-static perturbations. In fact, in the Newtonian regime the only signal of the dark sector interaction in structure formation to linear order is via the modification of the background expansion history.

The energy conservation and the Euler equation for density and velocity perturbations in the dark matter and the Poisson equation are given by
\begin{eqnarray}
\dot\delta_c+\frac{\theta_c}{a}&=&0 \,,\\
\dot \theta_c+H\theta_c+\frac{k^2}{a}\Phi&=&0 \,,\\
k^2\Phi&=&4\pi G_Na^2(\rho_c\delta_c + \rho_b\delta_b) \,.
\end{eqnarray}
These equations have precisely the same form as those for the
non-interacting case.

However, a non-trivial $Q$ is induced by the non-adiabatic scaling of $\rho_c$ such that $\rho_m\equiv\rho_c+\rho_b \neq \rho_m^s$ although $\rho_b=\rho_b^\s$.
The Poisson equation is written as
\begin{equation}
k^2\Phi=4\pi G_Na^2(\rho_b+\rho_c) \delta_m \,,
\end{equation}
where we used the fact that $\delta_m\equiv\delta_b=\delta_c$.
Comparing this with the definition of $Q$ in Eq.~(\ref{def:Q})
\begin{equation}
k^2\Phi=4\pi G_N a^2Q(a,k)\rho_m^\s\delta_m \,,
\end{equation}
we derive
\begin{equation}
\label{qc}
Q(a,k)=\frac{\rho_b+\rho_c}{\rho_m^\s}\,.
\end{equation}
In Fig. \ref{qcx} we plot the parameter $Q(a,k)$ from Eq.~(\ref{qc}), with different values of the interaction rate $\Gamma_c$.

\subsection{Model II: $\mathcal{C}=\Gamma_\de\rho_\de$}

Next, we consider the interaction given by $\mathcal{C}=\Gamma_\de\rho_\de$. A covariant form of the interaction can be found in \cite{Amendola:1999er}. The Euler equation for the velocity perturbation is the same as in the previous model
\begin{equation}
\dot \theta_c+H\theta_c+\frac{k^2}{a}\Phi=0,
\end{equation}
but the energy conservation equation for the dark matter density contrast now acquires a source term due to the coupling with the dark energy \cite{CalderaCabral:2009ja,Koyama:2009gd}:
\begin{eqnarray}
\dot\delta_c+\frac{\theta_c}{a}=\Gamma_\de\frac{\rho_\de}
{\rho_c}\delta_c\,.
\end{eqnarray}
This fact leads to a linear bias between dark matter and baryon over-densities. The source term leads to a modified growth equation of dark matter:
\begin{eqnarray}
&& \hspace{-5mm} \ddot \delta_c+2H\left(1-\frac{\Gamma_\de}{H}
   \frac{\bar\rho_\de}{\bar\rho_c}\right) \dot \delta_c = 
 \nonumber \\
&& \hspace{5mm} 4\pi G\Bigg\{ \bar\rho_b\delta_b + \left[1+
   \frac{2}{3a}\frac{\Gamma_\de}{H}\frac{\bar\rho_\de}{\bar\rho_c}\times\right.
 \nonumber \\
&& \hspace{5mm} \left. \times
   \left\{2-3w+\frac{\Gamma_\de}{H}\left(1+\frac{\bar\rho_\de}{\bar\rho_c}
   \right)\right\}\right] \bar\rho_c\delta_c \Bigg\}
\,. \label{evolutionx}
\end{eqnarray}
In this model, there is an ambiguity in the definition of $Q$ due to the bias between baryon and dark matter over-densities. The baryon over-densities obey the same growth equation as that for matter perturbations in the sDE reference model
\be
  \ddot{\delta}_b +2H\dot\delta_b - 4\pi G(\rho_b \delta_b + \rho_c \delta_c)=0 \,.
\ee
Therefore we define the function $Q$ with respect to the baryons
\begin{equation}
k^2\Phi=4\pi G_N a^2Q(a,k)\rho_m^s \delta_b \,,
\end{equation}
such that the bias is absorbed into $Q$. Hence we have
\begin{equation}
Q(a,k)=\frac{\rho_b\delta_b+\rho_c\delta_c}{\rho_m^s \delta_b} \,.
\label{qx}
\end{equation}
The reason why we chose to define $Q$ in terms of $\delta_b$ is that we might be able to use the energy conservation equation of the baryons to measure the baryon density contrast from peculiar velocities of galaxies, which are likely to follow the baryon peculiar velocities. Dark matter over-densities are subject to bias and we would not be able to measure them directly. 
The above expression for the induced $Q$ in case II is also applicable to other kinds of interaction models, e.g.~the one proposed by Amendola \cite{Amendola:1999er} where the interaction is described by a quintessence field coupled to dark matter. In Fig.~\ref{qcx} we plot the parameter $Q(a,k)$ from Eq.~(\ref{qx}) for different values of the interaction rate $\Gamma_\de$.

\section{Trajectories on the $\Sigma$ and $\mu$ plane}

In the previous three sections we examined predictions for $Q$ and $\eta$ in MG models and GR models with exotic dark energy. As discussed in section III, MG models characterised by the BD theory in the quasi-static approximation have a distinct path on the plane of ($\Sigma$, $\mu$). In this section, we summarise our predictions in this plane. For simplicity we only consider the case where $\Sigma$ and $\mu$ are scale independent.

\begin{figure}[t]
 \begin{center}
 \epsfysize=3.truein
   \epsffile{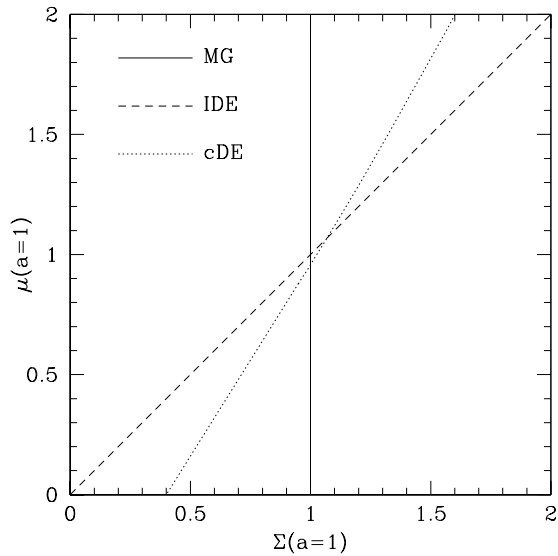}
   \caption{\footnotesize Trajectories on $\Sigma$ and $\mu$ plane of BD type MG models (solid curve), clumping dark energy (dotted curve) and interacting dark energy (dash curve).}
\label{fig:sigmamu}
\end{center}
\end{figure}

Let us consider the simplest massless case in the MG models described by the BD theory. In this case, as is seen from Eq.~(\ref{eq:MGqeta}), $Q$ and $\eta$ is expressed in terms of a single parameter $w_{BD}$, i.e.~both are not independent quantities.
If we project this constraint on the $\Sigma$ and $\mu$ plane, we get
\begin{equation}
\Sigma(k,a)=1, \quad \mu(k,a) = \frac{2(2+\om)}{3 + 2\om}\,,
\end{equation}
and interestingly, $\Sigma(k,a)=1$ is the same as GR with varying $\mu$ only. Note
that $\mu$ and $\Sigma$ are scale independent, as the massless case is considered. As is shown in Fig.~\ref{fig:sigmamu}, the MG models described by the BD theory trace a path along the direction of $\mu$ at fixed $\Sigma=1$.

In clustering dark energy (cDE) models, it is difficult to give physical models for dark energy perturbations. We have considered a simple toy model where the pressure perturbation is modeled as in Eq.~(\ref{dpressure}). If we impose scale-independence of $\mu$ and $\Sigma$, we find $g_p=0$ and that the anisotropic stress should be related to the pressure perturbation via $\sigma_\de = \delta{}P_\de/(1+w_\de) \rho_\de$. Then we are left with only one free parameter, either the pressure perturbation $\delta P_\de$ or the anisotropic stress $\sigma_\de$, thus again there is a unique path on the ($\Sigma$, $\mu$) plane. Of course, it is possible to vary $\delta{}P_\de$ and $\sigma_\de$ fully independently, but then the scale independence of $\Sigma$ and $\mu$ is lost, or absolute fine-tuning.

In interacting dark energy (IDE) models, no an\-i\-so\-trop\-ic stress is introduced with the assumption that dark energy is smooth and, thus, only $Q$ is non-trivial. Therefore, the condition $\eta=1$ reduces the degrees of freedom again to only one such that we have the relation
\be
\Sigma=\mu \,.
\ee
This turns out to be quite distinct from the trajectory of the MG models as is shown in Fig.~\ref{fig:sigmamu}.

These examples suggest that if we can measure $\Sigma$ and $\mu$ from observations,
the path on the ($\Sigma$, $\mu$) plane enables us to identify the underlying physics of the cosmic acceleration.

\section{Conclusions}

In this paper, we proposed to parameterise the relation between the lensing potential and the matter over-densities, $\Sigma$, and the dynamic relation between the Newtonian potential and the matter over-densities, $\mu$, which enable us to characterise theoretical models and constrain them with observations. If dark energy is described by a perfect fluid that is homogeneous (smooth) on sub-horizon scales these parameters are trivial, i.e.~$\Sigma=1$ and $\mu=1$. We showed that $\Sigma$ and $\mu$ can depart from unity in some theoretical models; such as modified gravity models, interacting dark energy models and clustering dark energy models. Interestingly, both parameters are related to each other in an unique way depending on the underlying theory:

\begin{itemize}
\item
With the assumption of the scale-independent evolution of perturbations, $\Sigma$ and $\mu$ in Brans-Dicke type MG models are described by a single variable, $\om$, which leads to a specific trajectory with $\Sigma=1$. This comes from the fact that there is no coupling between photons and the BD scalar field.

\item
In clustering dark energy the scale-independence of $\Sigma$ and $\mu$ and the simple assumption that $\delta P_\de\propto\delta\rho_\de$ lead to a constraint equation between the pressure perturbation and anisotropy stress. Therefore, both $\Sigma$ and $\mu$ are determined by a single variable, and consequently there is a unique trajectory in the ($\Sigma$, $\mu$) plane.

\item
For interacting dark energy models, as there is no anisotropic stress induced by interactions, we have a simple relation, $\Sigma=\mu$. Nevertheless, a non-trivial $\Sigma$ is induced by the non-adiabatic scaling of the background dark matter density.
\end{itemize}

Those investigations enrich our understanding of large scale structure formation, and provide theoretical priors on the modified growth parameter space spanned by $\Sigma$ and $\mu$.

\begin{acknowledgments}
Y-S.S.~is supported by the UK's Science \& Technology
Facilities Council (STFC). K.K.~is supported by the European Research
Council, Research Councils UK and STFC. L.H. is supported by the Swiss National Science Foundation (SNSF).
\end{acknowledgments}


\end{document}